\documentclass[sigconf]{acmart}
\usepackage{subcaption}

\AtBeginDocument{%
  \providecommand\BibTeX{{%
    \normalfont B\kern-0.5em{\scshape i\kern-0.25em b}\kern-0.8em\TeX}}}

\copyrightyear{2024}
\acmYear{2024}
\setcopyright{rightsretained}

\acmConference[e-Energy '24]{The 15th ACM International Conference on Future
and Sustainable Energy Systems}{June 4--7, 2024}{Singapore, Singapore}
\acmDOI{10.1145/3632775.3661959}
\acmISBN{979-8-4007-0480-2/24/06}
\begin{document}

\title{Exploding AI Power Use: an Opportunity to Rethink Grid Planning and Management}












\author{Liuzixuan Lin}
\affiliation{
  \institution{University of Chicago}
  \city{Chicago, IL}
  \country{USA}}
\email{lzixuan@uchicago.edu}

\author{Rajini Wijayawardana}
\affiliation{
  \institution{University of Chicago}
  \city{Chicago, IL}
  \country{USA}}
\email{rajini@uchicago.edu}

\author{Varsha Rao}
\affiliation{
  \institution{University of Chicago}
  \city{Chicago, IL}
  \country{USA}}
\email{varsharao@uchicago.edu}

\author{Hai Nguyen}
\affiliation{
  \institution{University of Chicago}
  \city{Chicago, IL}
  \country{USA}}
\email{ndhai@cs.uchicago.edu}

\author{Wedan Emmanuel Gnibga}
\affiliation{
  \institution{Univ. Rennes, Inria, CNRS, IRISA}
  \city{Rennes}
  \country{France}}
\email{wedan-emmanuel.gnibga@irisa.fr}

\author{Andrew A. Chien}
\affiliation{
  \institution{University of Chicago \& Argonne National Laboratory}
  \city{Chicago, IL}
  \country{USA}}
\email{aachien@uchicago.edu}

\renewcommand{\shortauthors}{L. Lin, R. Wijayawardana, V. Rao, H. Nguyen, W. E. Gnibga, A. A. Chien}

\begin{abstract}
The unprecedented rapid growth of computing demand for AI is projected to increase global annual datacenter (DC) growth from 7.2\% to 11.3\%. We project the 5-year AI DC demand for several power grids and assess whether they will allow desired AI growth (resource adequacy). If not, several ``desperate measures''---grid policies that enable more load growth and maintain grid reliability by sacrificing new DC reliability are considered.  

We find that two DC hotspots---EirGrid (Ireland) and Dominion (US)---will have difficulty accommodating new DCs needed by the AI growth. In EirGrid, relaxing new DC reliability guarantees increases the power available to 1.6x--4.1x while maintaining 99.6\% actual power availability for the new DCs, sufficient for the 5-year AI demand. In Dominion, relaxing reliability guarantees increases available DC capacity similarly (1.5x--4.6x) but not enough for the 5-year AI demand. New DCs only receive 89\% power availability. Study of other US power grids---SPP, CAISO, ERCOT---shows that sufficient capacity exists for the projected AI load growth.

Our results suggest the need to rethink adequacy assessment and also grid planning and management.  New research opportunities include coordinated planning, reliability models that incorporate load flexibility, and adaptive load abstractions.
  
\end{abstract}

\ccsdesc[500]{Applied computing~Data centers}
\ccsdesc[500]{Hardware~Power and energy}
\ccsdesc[300]{Computing methodologies~Artificial intelligence}

\keywords{Power grid, Data centers, Resource adequacy, AI}



\maketitle

\section{Introduction}
\label{sec:intro}

The phenomenon of ChatGPT has led to an explosion of interest and use of generative AI chatbots and intelligent assistant services based on large language models (LLMs).  
LLMs have been incorporated into many widely used computing services such as Slack, Microsoft Office and stimulated new tools like Copilot for programming, etc. Training and operating these large models is costly \cite{genAICost, carbon-generative-ai}, and the ``AI gold rush'' has triggered massive investment in GPU hardware and datacenters \cite{goldmanSachs2023, meta-nvidia-shipment}. For example, 
the three largest hyperscalers (Amazon, Microsoft, and Google) reported large increases in datacenter capex investment from \$78B (2022) to \$120B (2023), a \$42B or 54\% increase \cite{Econ23}. 
There is increasing awareness of  exponential datacenter growth due to AI and resulting energy crisis \cite{economist-ai-dc24,de2023growing,bloomberg_ai_power2024,verge_ai_demands2024,semi-analysis-ai-power,AI-disrupt-green-dc-efforts,dc-power-surge-NYtimes,aventine-ai-power}: the existing and planned power grid infrastructure cannot support it, leading to 
climate-unfriendly measures (delaying  shutdown of fossil-based generation). However, this awareness focuses on total power load growth, not specific grid  challenges. 

Datacenters' rapid load growth is outracing both annual power grid planning cycles and investment/construction cycles (5--20 years) \cite{eirgrid22report,dominionVA23report,pjm23loadForecast,dc-power-demand-surge}.
A number of power grids have halted or slowed new datacenter connections 
to ensure grid reliability \cite{NoVAHaltDC,IrelandHaltDC}.  For power grids, the situation is compounded by additional stresses from extreme climate events and increased generation volatility from increasing dependence on renewables \cite{ChinaGridOutage,USGridOutage,LongTermCost20}. 
Therefore, to better understand the challenges and explore new research opportunities, it is necessary to assess the demand growth against \textit{grid resource adequacy}, the method by which power grids decide how much load can be safely connected to the grid.


In this paper, we study a diverse set of power grids---EirGrid (Ireland), Dominion, CAISO, ERCOT, and SPP (US), assessing whether they have enough reliable
capacity to meet the 5-year AI/Cloud datacenter power growth.
We examine how relaxing new datacenter power reliability guarantees could increase the available grid power for new datacenters.
Specific insights include:

\begin{itemize}
\item In EirGrid, we project that AI will increase the DC compound annual growth rate (CAGR) to 16.7\%.  Its resource plan cannot accommodate the projected load in 2024.  Relaxing power reliability for new datacenters increases the power available for new datacenters to 1.6x--4.1x so the AI demand can be met for the next 5 years. New datacenters can expect 99.6\% power availability despite lower guarantees.

    
\item Dominion grid, which faces a similar rapid DC growth challenge (23.6\% CAGR), cannot meet the growing demand even with reduced new datacenter power reliability.  The new datacenters will experience poor power availability ($<$ 90\%).

\item Study of several other power grids (CAISO, ERCOT, and SPP) finds that they have excess capacity to accommodate AI load growth within reliable grid operation. 

\item We identify new research opportunities to rethink adequacy assessment, 
reinvent grid planning and management as cooperative, design new reliability models that incorporate
load flexibility, and adaptive load abstractions.
\end{itemize}




\section{Problem}
\label{sec:problem}



\begin{figure}[h]
    \centering 
    \includegraphics[width=\columnwidth]{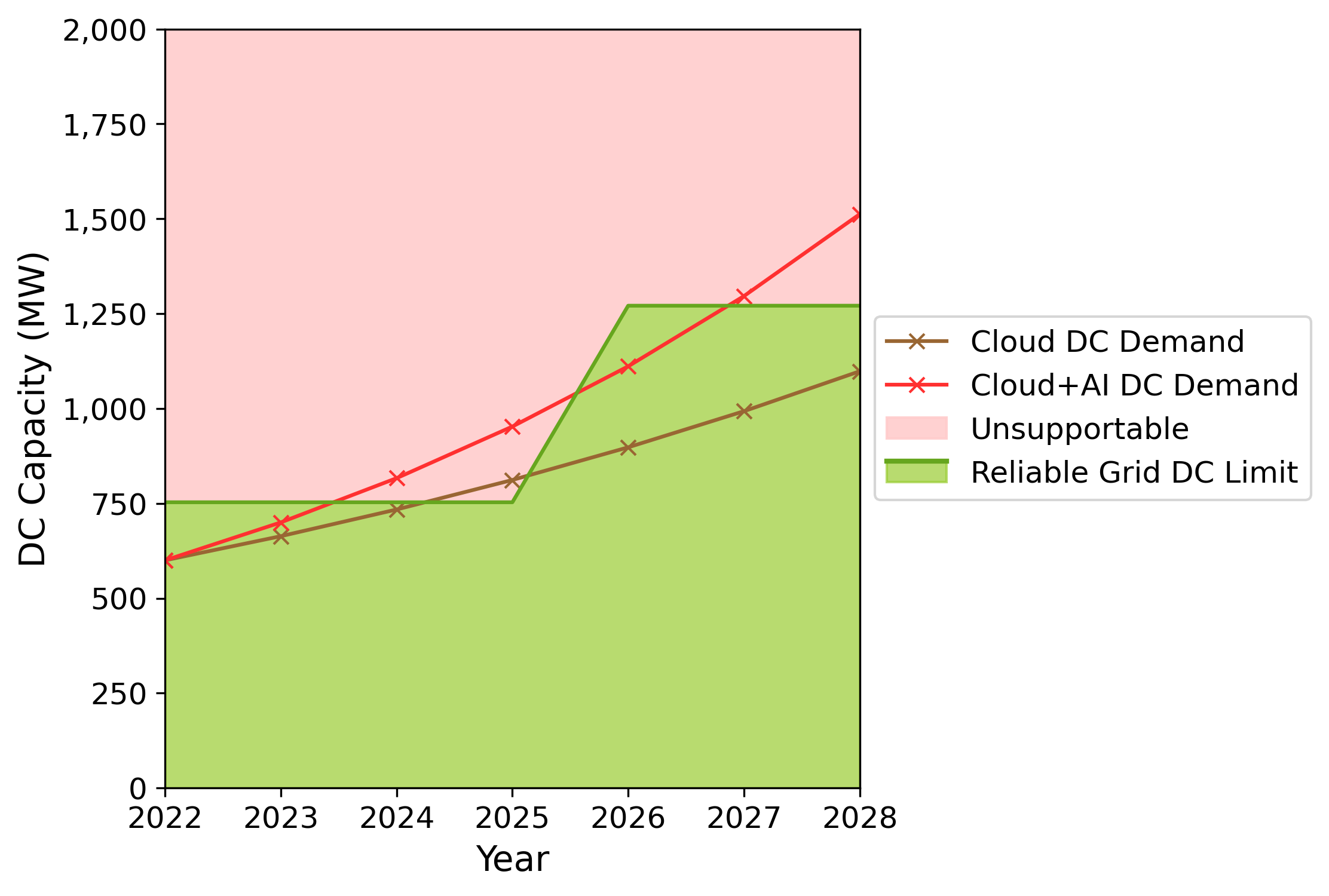}
    \caption{EirGrid DC load is growing rapidly, and AI accelerates it.  By 2025, Cloud DC growth cannot be supported, and by 2024,  Cloud+AI DC growth cannot be supported.} 
    \label{fig:eirgrid_admissionVSdemand}
\end{figure}

Consider EirGrid, a hotspot for datacenters with 600 MW datacenter load (14\% of grid load) in 2021 \cite{eirgrid22report}.
First, we project Cloud growth and Cloud+AI growth using the methodology in Section \ref{sec:load-growth-model}.
For Cloud, EirGrid's plan expected a 10.6\% CAGR (compound annual growth rate) DC growth (brown line), and AI projections increase 
the growth rate to 16.7\% (red line) as shown in Figure \ref{fig:eirgrid_admissionVSdemand}.

Using the grid resource adequacy assessment methodology in Section \ref{subsec:methodology}, we compute the reliable grid limit for 2023-2028 (see \cite{eirgrid22report}).  The results show that 
for Cloud, demand will exceed the reliable grid limit by 2025, and with the more rapid growth of Cloud+AI, demand will exceed the limit even earlier, by 2024.



\begin{figure}[h]
    \centering 
    \includegraphics[width=\columnwidth]{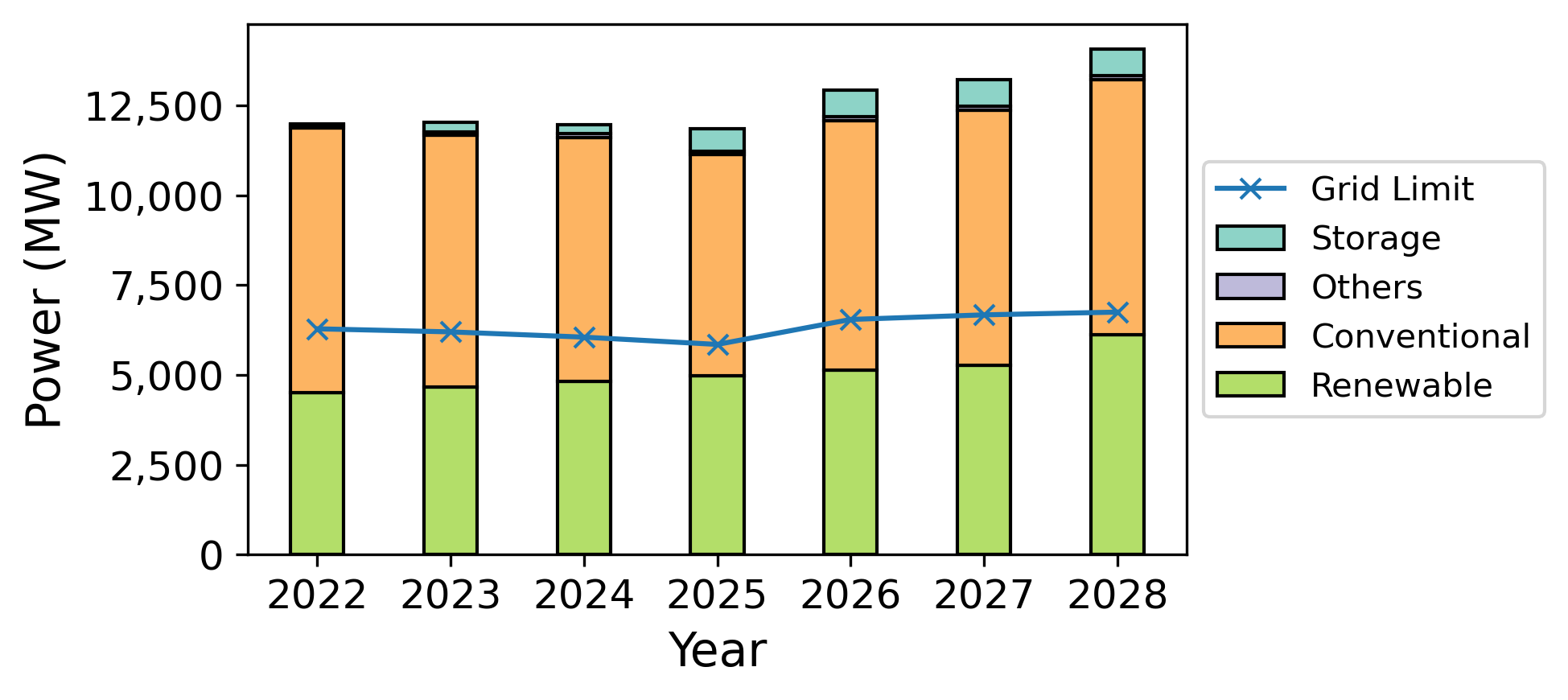}
    \caption{EirGrid's Reliable Load Limit (blue line) and Resources, 2022--2028. The load limit does not increase despite a 30\% increase in renewable generation capacity.}
    \label{fig:eirgrid_resource}
\end{figure}

The shape of the reliable DC limit is determined by changes in 
generators available on the EirGrid over time (Figure \ref{fig:eirgrid_resource}).
From 2023 to 2026, EirGrid plans to replace 1,500 MW of conventional generation with renewables.  This maintains total power available, but increases volatility.  Consequently,  the reliable load limit decreases, and cannot support the increased datacenter demand.  

In 2026, the addition of 800 MW of gas generation lifts the reliable grid limit.  A further planned addition of 1,000 MW wind generation in 2028 increases the reliable load limit only marginally.  Thus capacity available for new DC load is reduced by substitution of renewables (for conventional generation), and only partially met by new gas generation.
To conclude, EirGrid will not be able to support growing DC load by as early as 2024, and certainly by 2025.





\section{Study: EirGrid}
\label{sec:eirgrid}

We describe the resource adequacy (RA) assessment methodology for EirGrid.  Then, we consider two different approaches 
to increase the supportable datacenter load without endangering power grid reliability.  These methods require no physical changes, and thus can be applied immediately.  Using these methods, we show how they could accommodate growing Cloud+AI datacenter load, and achieve high datacenter availability.


\subsection{Methodology}
\label{subsec:methodology}

\subsubsection{Resource Adequacy Assessment}
\label{subsubsec:resourceAdequacyAssessment}
We use a grid resource adequacy assessment framework based on the published methodologies of a variety of power grids \cite{eirgrid22report,dominionVA23report,caisoPlan}.
With grid current and planned (2022--2028) generation resources (conventional and renewables, energy storage, and others such as imports), 
load and renewable generation traces are used to assess whether grid load can be met for each time interval in a year. Simulating over a year captures the day-to-day variations in load and renewable generation, and their relationship. To model future grid scenarios, 2022's non-DC load and renewable generation traces are scaled proportionally based on expected load growth and planned generation capacity, respectively.  The result of resource adequacy assessment is the \textbf{loss of load expectation (LOLE)} metric, whose definition and standard vary by grid.  For EirGrid, LOLE is the expected number of hours in a year when the available grid capacity cannot meet load, and the target is $\le$ 8 hours per year (99.9\% reliability).  Formally, for year $y$:

\begin{equation}
    LOLE_y=(\sum_{t=1}^T I_{y,t})*intLen/60
\end{equation}
where $T$ is the number of data time intervals in a year and $intLen$ is the length of time interval in minutes (15 minutes for EirGrid). $I_t$ indicates whether capacity shortage happens at time $t$:
\begin{equation}
\label{eq:powerBalance}
    I_{y,t}=
    \begin{cases}
    1,&load_{y,t}>renewGen_{y,t}+convCap_{y}+stoCap_{y}+other_{y}\\
    0,&otherwise
    \end{cases}
\end{equation}
\begin{equation}
    load_{y,t}=load_{DC,y,t}+load_{NonDC,y,t}
\end{equation}
where $renewGen_{y,t}$ is the actual renewable generation, and $convCap_{y}$, $stoCap_{y}$, and $other_{y}$ are capacities of conventional generation, energy storage, and other resources de-rated by availability factors (details in Appendix \ref{appendix:GridRADetails}). The grid datacenter limit is defined as the maximum datacenter load that produces LOLE $\le$ 8 hours.

The major uncertainties that this method doesn't model are year-to-year weather changes and load profile reshaping with load flexibility.  These are both opportunities for future research (see Section \ref{sec:research-oppty}).

\subsubsection{Datacenter Load Growth Model}
\label{sec:load-growth-model}

Projected Cloud datacenter power demand is based on EirGrid's forecast of Cloud growth of 10.6\% CAGR \cite{eirgrid22report}.  Cloud+AI demand adds the incremental AI demand estimated by extrapolating NVIDIA's latest datacenter revenue guidance released in August 2023 \cite{NVIDIAAugPressResease}.  This AI demand estimation produces 1,400 MW additional global datacenter capacity (4.1\% of current capacity) next year (details in Appendix \ref{appendix:loadGrowth}).   Because datacenter growth is not uniform geographically, we make a location-based adjustment that assumes more new capacity will be deployed in higher-value regions (e.g. EirGrid, Dominion) based on growth rates for Cloud alone.  For region $r$,
$$CAGR_{cloud+AI,r}=CAGR_{cloud,r}+\frac{CAGR_{cloud,r}}{CAGR_{cloud,avg}}\cdot \Delta CAGR_{AI,avg}$$
with $CAGR_{cloud,avg}$ (global average) of 7.2\%, the datacenter demand growth including AI comes to 16.7\% CAGR for EirGrid.

\subsubsection{New Schemes to Accommodate Datacenter Load} 
\label{subsubsec:DC-connection-schemes}

To avoid threatening grid reliability, researchers and,  more recently, commercial players have proposed unreliable power service for datacenters \cite{ieDCConnMeasures,lfltf,YangChien-TPDS17}, in effect, reducing the quality-of-service (QoS) for datacenter power from reliable to partially reliable.  So here, we consider datacenter grid connection with reduced reliability, and examine what the consequences are for the grid and the datacenters---outages.
\textbf{Datacenter outage} is defined as time periods when datacenter load is shed by the grid, and we report the outage rate. 


We study three different levels of QoS for new datacenters:

\begin{itemize}
    \item \textbf{Reliable.} The usual assumption in grid resource adequacy assessment \cite{eirgrid22report}.  Datacenters get reliable, continuous power supply up to maximum capacity.
    
    \item \textbf{80\% Reliable. } Datacenters get reliable, continuous power supply up to 80\% of the capacity.  The 20\% is only available when it doesn't tax the grid (see \cite{ieDCConnMeasures}).

    \item \textbf{0\% Reliable. } 
    Datacenters get reliable, continuous power supply for 0\% of the capacity.  That is, no guaranteed power.  The 100\% is only available when it doesn't tax the grid.  The outage rate is limited to 1\% (88 hours in a year).
An example is ERCOT's Large Flexible Loads program \cite{lfltf} or the ideas espoused in Zero-carbon Cloud \cite{YangChien-TPDS17}.

\end{itemize}

The benefit of reducing QoS for new datacenters is more DC load can be safely attached to the grid.  If the grid is overloaded, the DC load can be reduced, protecting grid reliability while avoiding grid service violations as these reductions do not count as LOLE.




\subsection{Results}

\begin{figure}[h]
    \centering 
    \includegraphics[width=\columnwidth]{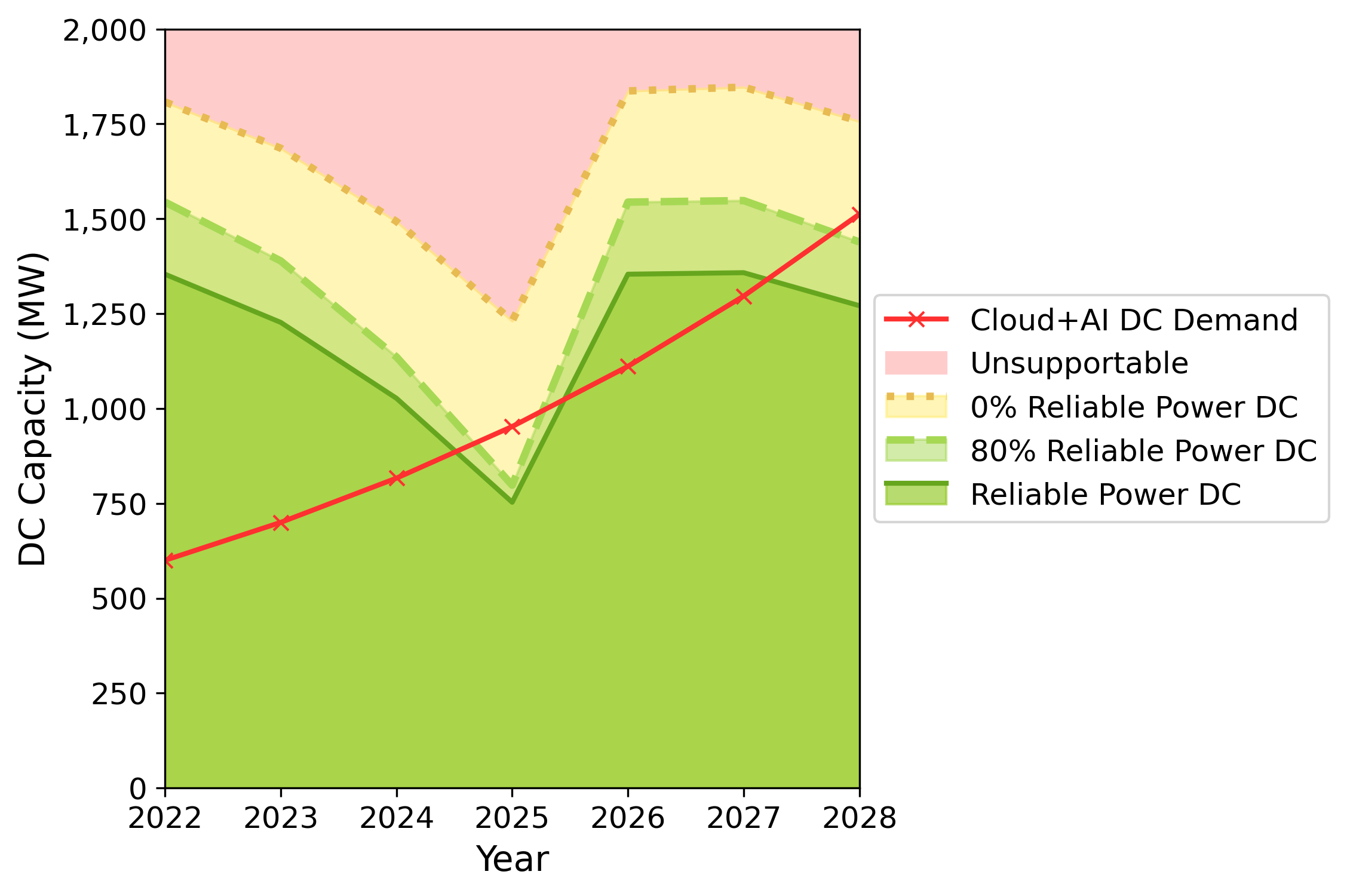}
    \caption{EirGrid Grid Datacenter Capacity Limits with Varied New Datacenter Power QoS vs. Demand.}
    \label{fig:eirgrid_supportableByYear}
\end{figure}

We compute resource adequacy limits under the reduced DC QoS schemes, and plot the results in Figure~\ref{fig:eirgrid_supportableByYear}.  
Note that Figure \ref{fig:eirgrid_supportableByYear} shows annual capacity limits, differing from Figure \ref{fig:eirgrid_admissionVSdemand} which shows stable capacity available into the future (constrained by future minimum). 

Relaxing new DC QoS to 80\% reliable enables EirGrid to accommodate 25\% greater datacenter capacity, a significant improvement, but still insufficient to support the increasing AI demand.  
Further reducing new DC QoS to 0\% reliable increases the available new datacenter capacity to 1.6x--4.1x, enabling the EirGrid to support ALL of the projected Cloud+AI demand through 2028.  Note that the demand increase from 2023 is more than 100\% (2x).    

\begin{figure}[h]
    \centering 
    \includegraphics[width=0.9\columnwidth]{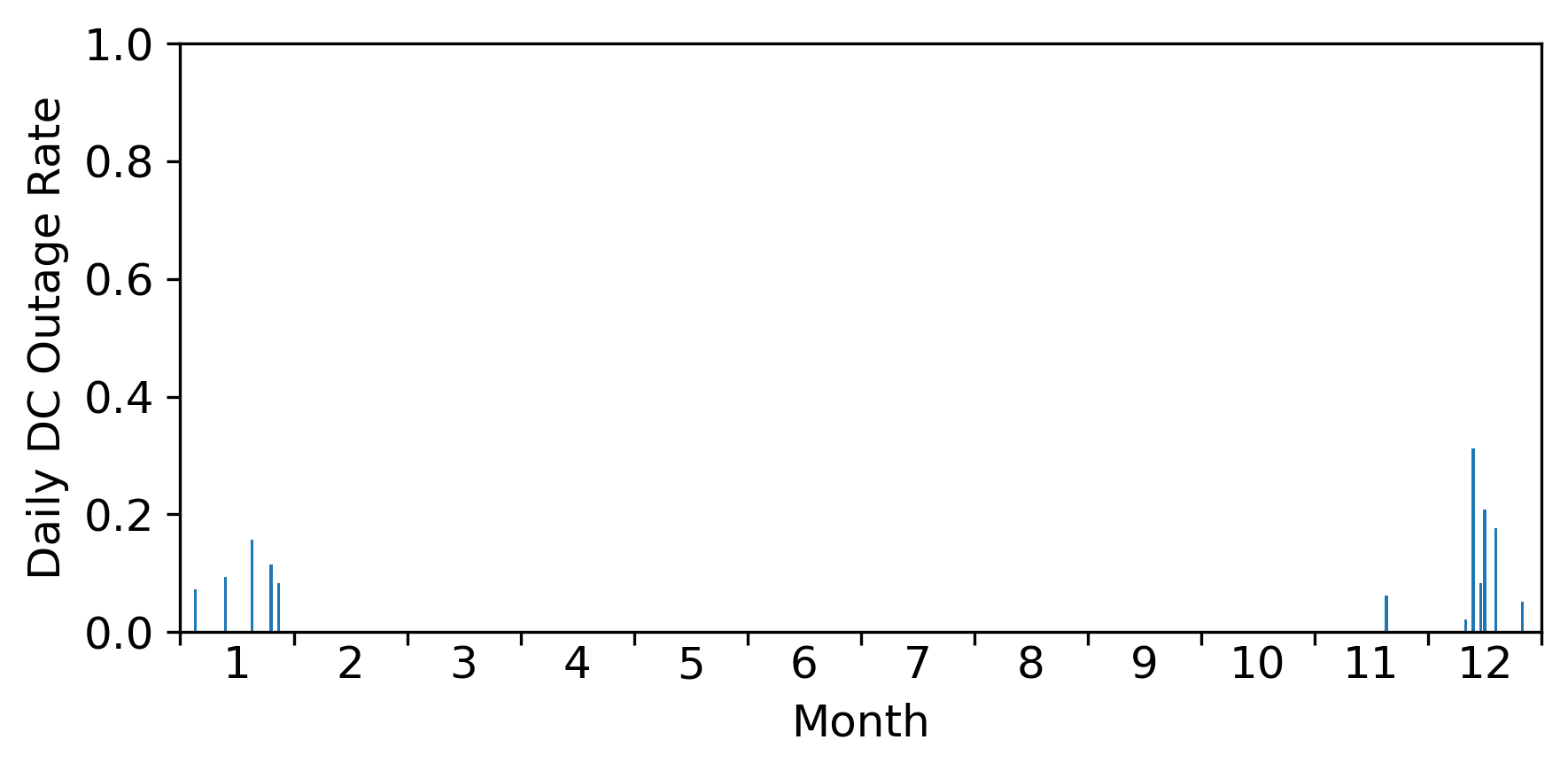}
    \caption{Datacenter Outage Rates (daily fraction),
    2028 EirGrid: using 0\% reliable DCs to meet 900 MW AI demand. Outages are rare (0.4\%, 34.5 hours) and seasonal (winter).}
    \label{fig:eirgrid_dcAvailability}
\end{figure}

How often does the grid fail to fully power the reduced QoS datacenters?
We computed the 
EirGrid daily datacenter outage rate for 2028 with 0\% reliable new datacenter QoS (Figure~\ref{fig:eirgrid_dcAvailability}).  Each bar reports the daily fraction of time when there is datacenter load shedding.
Shedding occurs primarily in winter (beginning and end of year) and in total is 34.5 hours in a year, and typically a small magnitude.  So despite no guarantee, the new datacenters actually experience power supply QoS of 99.6\%, nearly the grid LOLE goal of 99.9\%.  We expect that with in-advance grid alerts \cite{eirgridAlert}, the negative impacts on datacenter operation should be minimal.  This suggests research on dynamic adequacy (dynamic models for reliability) is an interesting opportunity.

\section{Other Grids}
\label{sec:otherGrids}
We continue assessing whether the DC load growth can be met in the other four power grids (Dominion, CAISO, ERCOT, SPP). These grids vary in size, load, and generation mix, and have well-documented resource plan and datacenter capacity (Appendix \ref{appendix:GridRADetails}, \ref{appendix:loadGrowth}).

With proximity to the US government and transatlantic networking, Northern Virginia is home to the world’s highest concentration of data centers \cite{globalDC2023,Greenpeace-Nova19}, whose power demand is mainly served by Dominion Energy.   This datacenter concentration is expected to continue, and its growth has already exceeded grid operator's expectation several times, as reflected in Dominion's repeated revisions on load forecast \cite{dominionVA23report}.   In 2022, Dominion forecast 15\% CAGR datacenter load growth \cite{pjm23loadForecast} before 2028, and we project (Section \ref{sec:load-growth-model}) with AI growth this DC growth will rise to 23.6\% CAGR.

\begin{figure}[h]
    \centering
    \includegraphics[width=\columnwidth]{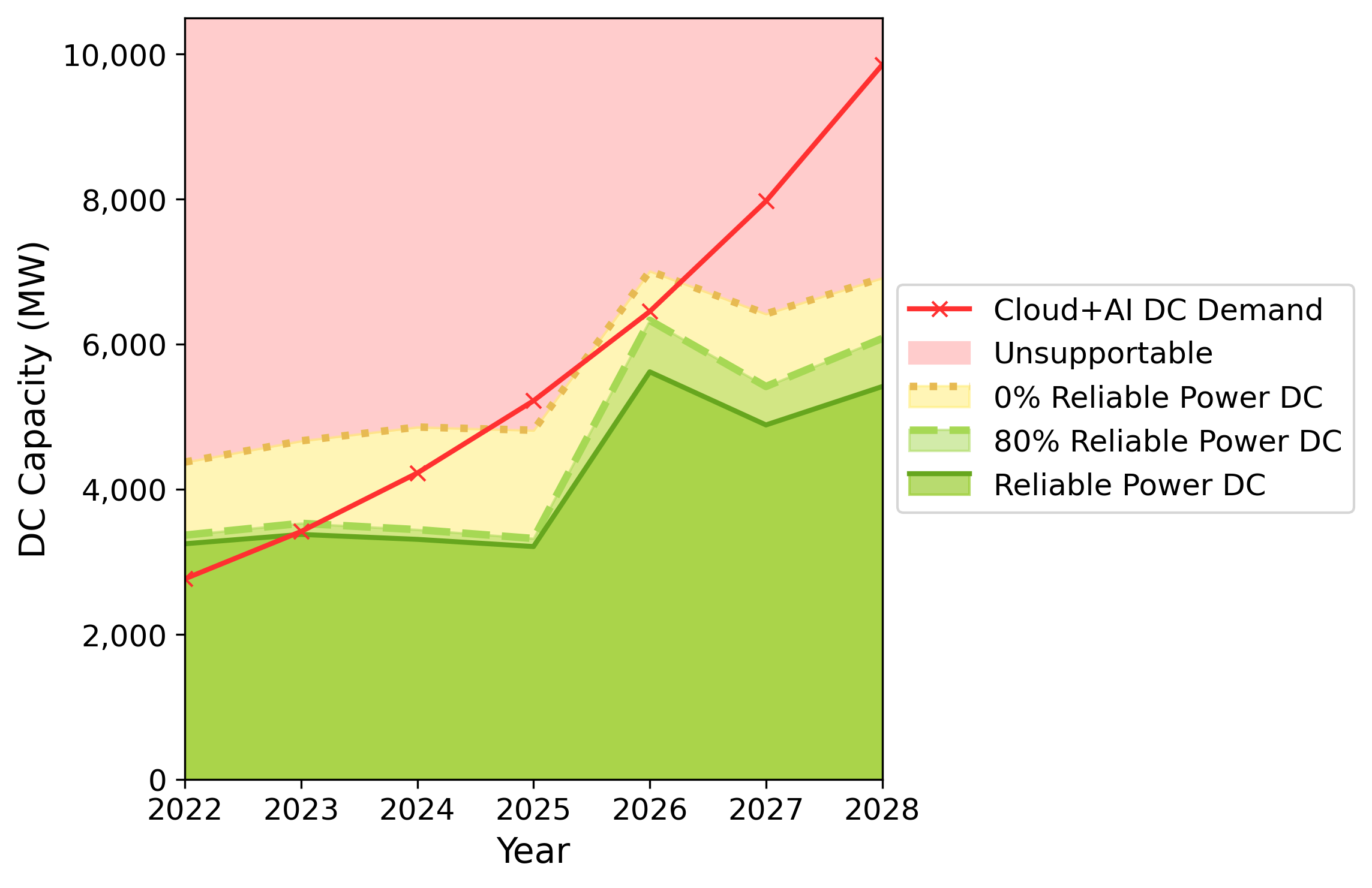}
    \caption{Dominion Grid Datacenter Capacity Limits with Varied New Datacenter Power QoS vs. Demand.}
    \label{fig:dominion}
\end{figure}

\begin{figure}[h]
    \centering 
    \includegraphics[width=0.9\columnwidth]{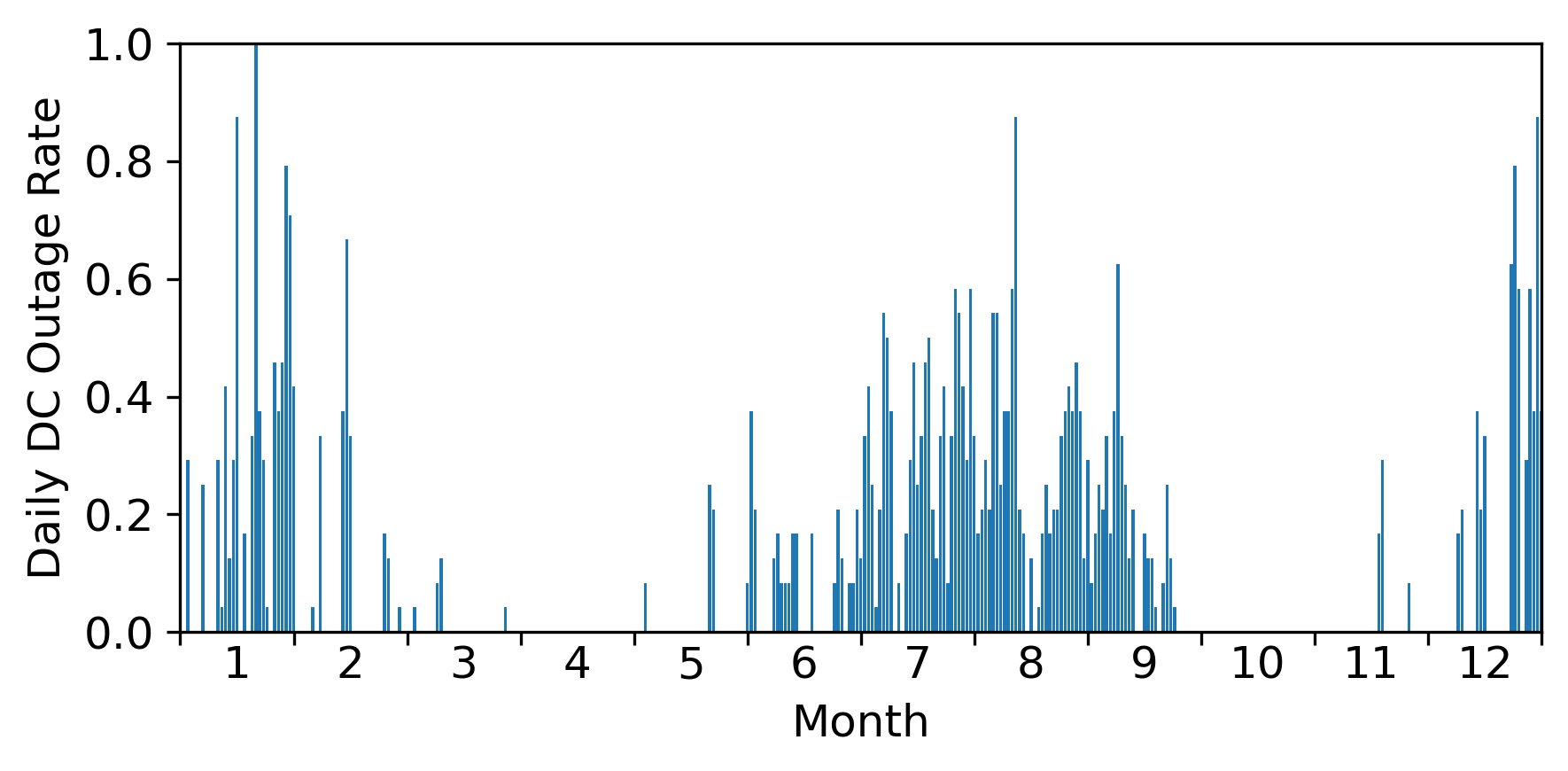}
    \caption{Outage Rates (daily fraction), 2028 Dominion Grid: using 0\% reliable datacenters to meet 7,100 MW AI-demand. Winter and Summer see 1,010 hours' outage in 145 days.}
    \label{fig:dominion_dcOutage}
\end{figure}

Figure \ref{fig:dominion} compares the grid datacenter capacity limits and the demand growth.  The projection shows that the existing 2,767 MW datacenter capacity \cite{dominionVA23report} will grow to 9,800 MW in 2028. However, the reliable datacenter limit under Dominion's planned resources \cite{dominionVA23report} is only 55\% of the demand.  Relaxing new datacenter QoS to 80\% reliable fails to meet DC demand.  Going further to  0\% reliable increases the limit but still only 70\% of the demand. With 100\% demand, datacenters added under 0\% guarantee (Figure \ref{fig:dominion_dcOutage}) will experience frequent outages (1,010 hours in a year, 11.5\% time) unacceptable to datacenter operators.  In summary, for Dominion, relaxing new DC QoS increases capacity, but cannot meet the growing AI DC load.  Actually, it already has shortfalls \cite{NoVAHaltDC}, and new research on cooperative capacity planning (and investment) is needed (see Section \ref{sec:research-oppty}). 



\begin{figure}[h]
    \centering 
    \includegraphics[width=\columnwidth]{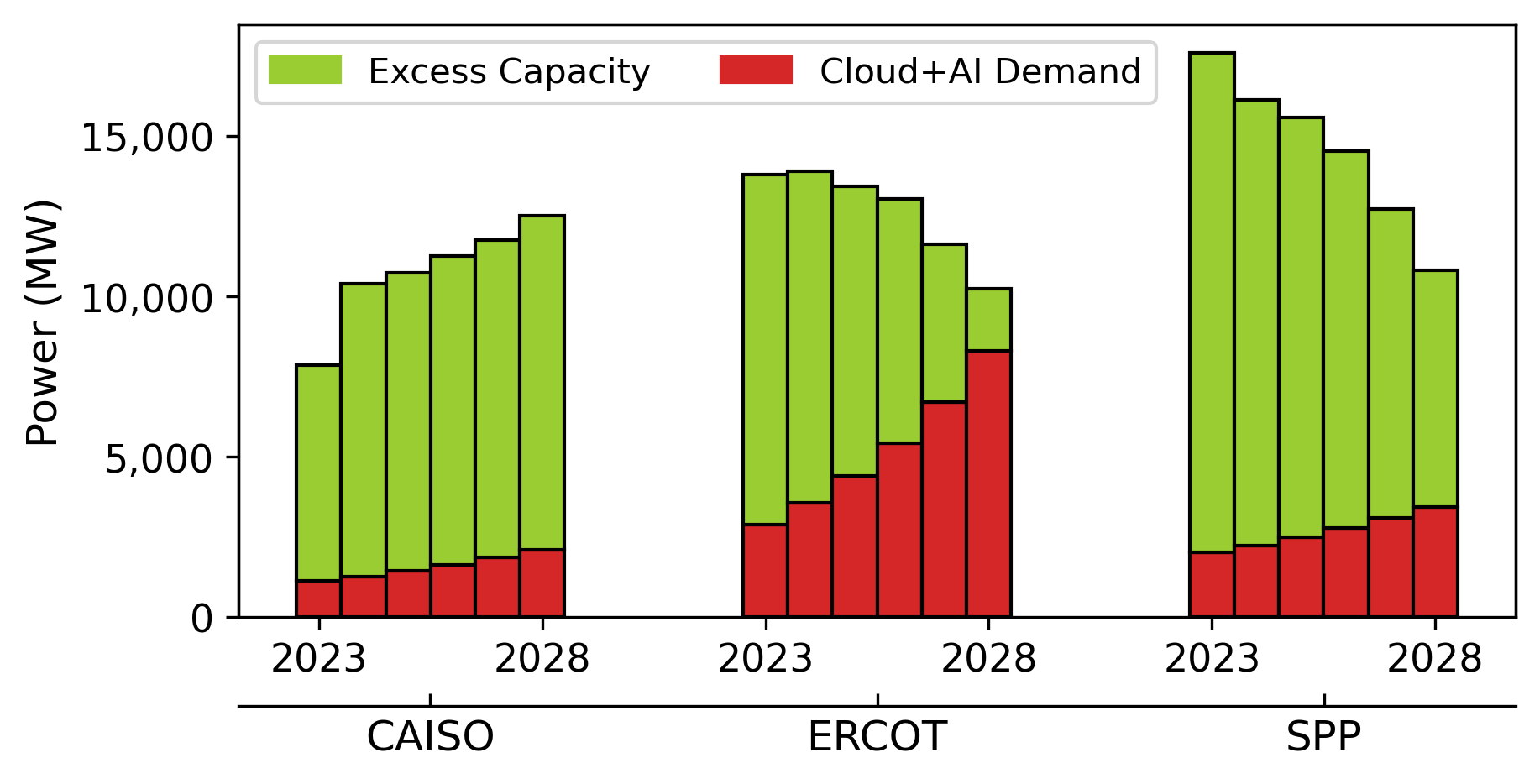}
    \caption{Excess Datacenter Capacity in CAISO, ERCOT, and SPP for 2023--2028, despite increasing Cloud and AI demand.}
    \label{fig:otherGrids}
\end{figure}

For the other three grids, excess datacenter capacity (available reliable capacity minus demand) will exist for the next 5 years (Figure \ref{fig:otherGrids}): new datacenters can get 100\% reliable power supply in these grids. The reasons for excess datacenter capacity vary by grid.

CAISO (California, US) has the most aggressive decarbonizaton plan, 
targeting 60\% RPS by 2030 and 90\% by 2035 \cite{caisoRPSGoal2035}. It plans to retire 6,000 MW conventional and add 12,000 MW solar and 9,000 MW wind by 2028 \cite{caisoPlan}. CAISO's strategy can bring 73\% \cite{caisoStorage} of the total 15,296 MW energy storage online for resource adequacy during peak hours. 

ERCOT (Texas, US) plans to add 30,000 MW solar and 11,000 MW energy storage by 2028 \cite{ercot2022study,ercot2023plan} to support energy demand growth of 23\% (industry growth) \cite{ercot2023loadForecast}.  Grid resources are adequate for datacenter growth if we use CAISO computation for energy storage contribution, but under ERCOT's pessimistic assumption (0 contribution during peak hours), it might not meet AI demand.  Further, as an isolated grid, it faces risk during extreme events \cite{ercotIsolationProblem}.

SPP (Southwest, US) has a large reserve margin (20\%) in generation resources as of 2023 and will maintain sufficient conventional generation balancing additions and retirements by 2028 \cite{spp2023plan}. This conservative plan meets projected demand.

\break
\section{Revisiting AI Power Growth and Grid Readiness}
\label{sec:acceleratingGrowth}
In the six months since we began this study in September 2023, concern about AI's increasing power  demand and the threat it represents to grid stability and decarbonization has become a headline concern \cite{dc-power-surge-NYtimes}.  In this section, we update our demand growth projection and grid readiness as of March 2024.

\begin{figure}[ht]
    \centering
    \begin{subfigure}[b]{0.23\textwidth} 
        \includegraphics[width=\textwidth]{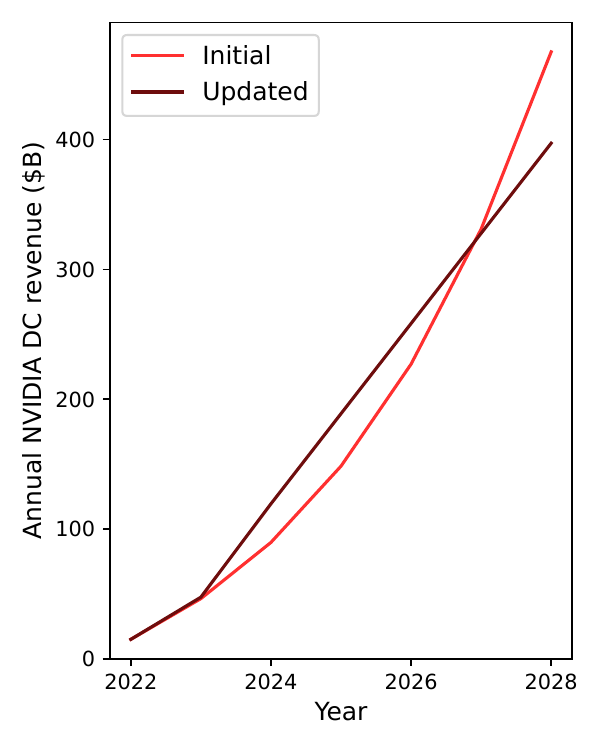}
        \caption{NVIDIA's Annual DC Revenue}
        \label{fig:updatedNvidiaRev}
    \end{subfigure}
    \hfill
    \begin{subfigure}[b]{0.23\textwidth} 
        \includegraphics[width=\textwidth]{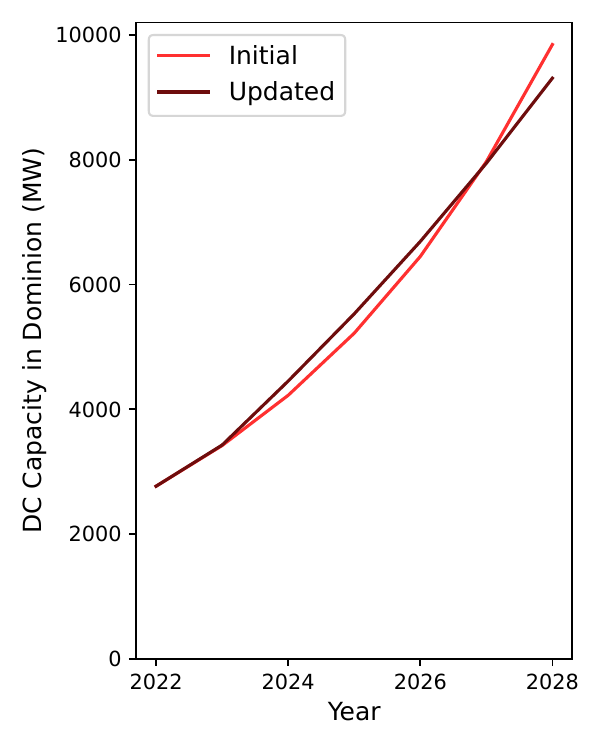}
        \caption{Dominion Grid DC Demand}
        \label{fig:updatedDominionLoad}
    \end{subfigure}
    \caption{Comparison between the Updated (linear) and Initial (exponential) Datacenter Projections.}
    \label{fig:updatedProjections}
\end{figure}

Continued growth in AI infrastructure \cite{saudiArabia40B,msftSpainAIInvestment},  and growth of NVIDIA's datacenter revenue indicate that AI power consumption growth is accelerating. Considering NVIDIA's latest actual and projected datacenter revenue (Appendix \ref{appendix:loadGrowth_update}), the four most recent quarters 
produce a 12-month total revenue of \$66.6B. The trend suggests linear datacenter revenue growth. Thus, we propose a linear DC  growth model, less aggressive than our initial model (exponential, Section \ref{sec:load-growth-model}) 
and evaluate its effects. 
Based on NVIDIA's average \$4.34B quarterly datacenter revenue increase, we project through 2028. The resulting annual datacenter revenue and corresponding power load are shown in Figure \ref{fig:updatedProjections}. Compared with our previous model, the new model reflects a slightly higher load in the near-term and a lower load in the long-term.  We consider this projection more realistic.

\begin{figure}[h]
    \centering 
    \includegraphics[width=\columnwidth]{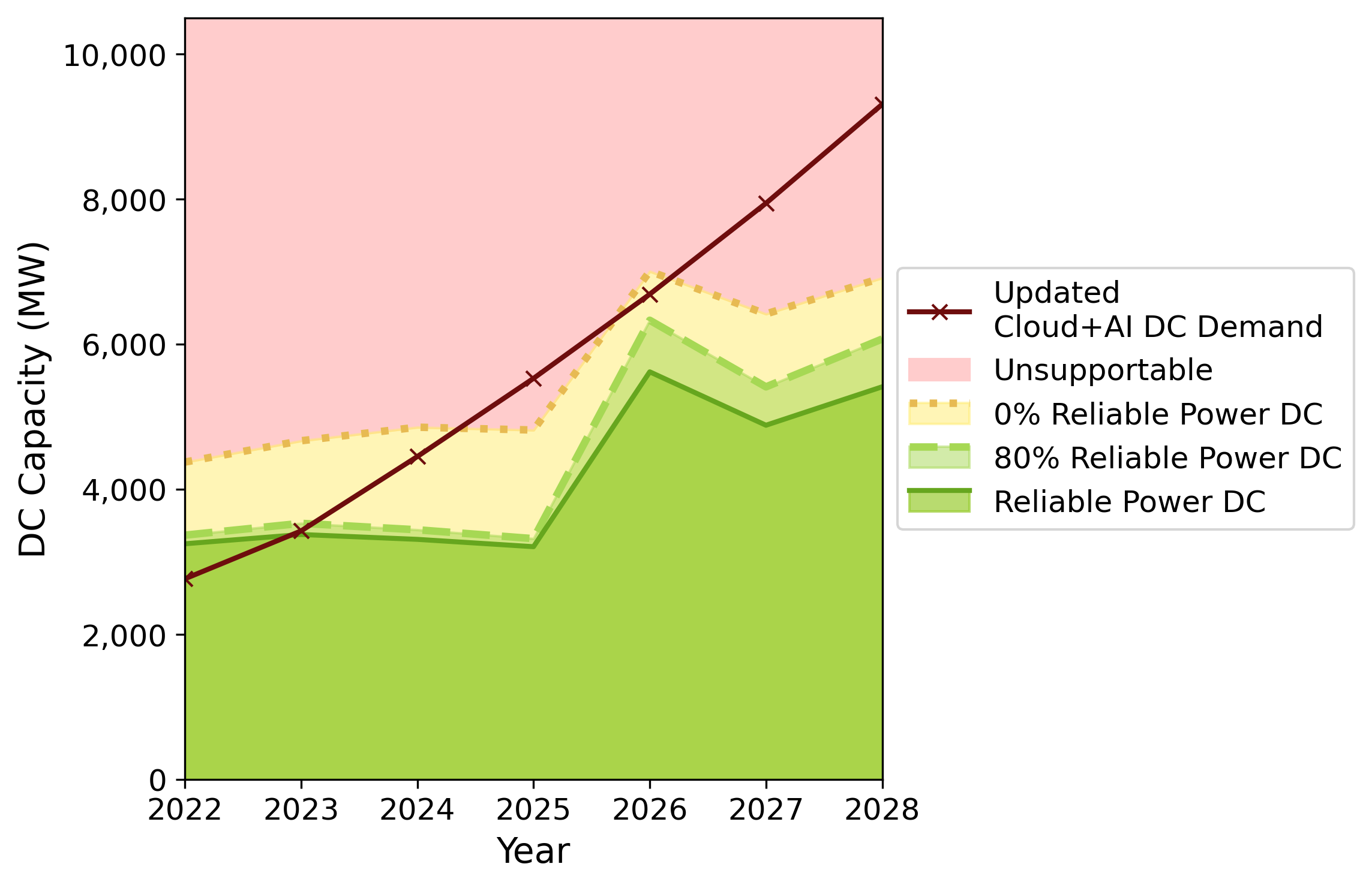}
    \caption{Dominion Grid Datacenter Capacity Limits vs. Updated Demand.}
    \label{fig:dominion_udpatedDemand}
\end{figure}

We compare the updated Dominion load growth projection to the capacity limits in 
Figure \ref{fig:dominion_udpatedDemand}.
Dominion has not released its 2024 Integrated Resource Plan, 
so we see higher demand against the same capacity limits. Our conclusions are essentially the same with updated demand:  shortage begins in 2024, and relaxing new datacenter QoS to 0\% reliable can still only meet 74\% of the demand in 2028. For the other four grids we studied, EirGrid added more renewable generation and energy storage resources in accordance with Irish government's updated goal of 80\% (was 70\%) renewable electricity by 2030 \cite{eirgrid23report}.  These resources can increase the reliable datacenter capacity limit in EirGrid to meet the demand. We have not found any changes in the other three US grid (CAISO, ERCOT, SPP) plans, perhaps because they are expected to have excess datacenter capacity in our assessment. However, rising awareness of energy demand growth challenges \cite{caiso2024tpp} suggests the situation may change in the future as resource adequacy concerns are spreading to more grids \cite{pjm24loadForecast}.

\section{Research Opportunities}
\label{sec:research-oppty}


The insight that the grid infrastructure cannot meet rapid Cloud+AI growth opens many research opportunities---rapid change is essential to solve this and other renewable grid problems.  We describe a few exciting new research opportunities below.


\paragraph{Coordinated Planning and Co-investment}
Traditional extrapolation 
\cite{dominionVA23report,eirgrid22report} is insufficient to track exponential growth
(e.g. the AI boom), see \cite{pjm23loadForecast,pjm24loadForecast}.  Now that datacenter loads are large, we need new information sharing and co-investment (de-risking) by wealthy computing companies to ensure sufficient power infrastructure for 3, 7, even 10-year windows.

\paragraph{New Models for Reliability}
Traditional statistical models for reliability were appropriate for loosely coupled, uncontrollable loads.  With large fluctuations in renewables and advanced load control beyond traditional demand-response \cite{lfltf}, how can we assess adequate supply and reliability?  Quantity of generation, ramping, etc. are not enough for tightly coupled, dynamically controlled systems.

\paragraph{New Abstractions for Load and Adaptive Load Management} Recent planning reports for renewable-dominated grids exploit terms such as ``Shape, Shift, Shed, and
Shimmy'' \cite{gerke2020california}, to describe load properties.  Should datacenters describe their load in these terms---to shape computing resource management and QoS for services? \cite{lin2012dynamic,YangChien-TPDS17,radovanovic2022carbon,xing2023carbon,zhang2021scheduling} Can this
enable cooperative datacenter and grid planning? \cite{le2016joint,lin2023adapting}

\begin{acks}
We thank the anonymous reviewers for the insightful reviews. This work is supported in part by NSF Grant CNS-1901466, and the VMware University Research Fund. We also thank the Large-scale Sustainable Systems Group members for their support of this work!
\end{acks}


\newpage
\bibliographystyle{ACM-Reference-Format}
\bibliography{bib/zccloud-10-2020, bib/powerLimit,bib/hotcarbon23}

\appendix
\section{Grid Resource Adequacy Assessment Details}
\label{appendix:GridRADetails}

\begin{figure}[h]
    \centering 
    \includegraphics[width=\columnwidth]{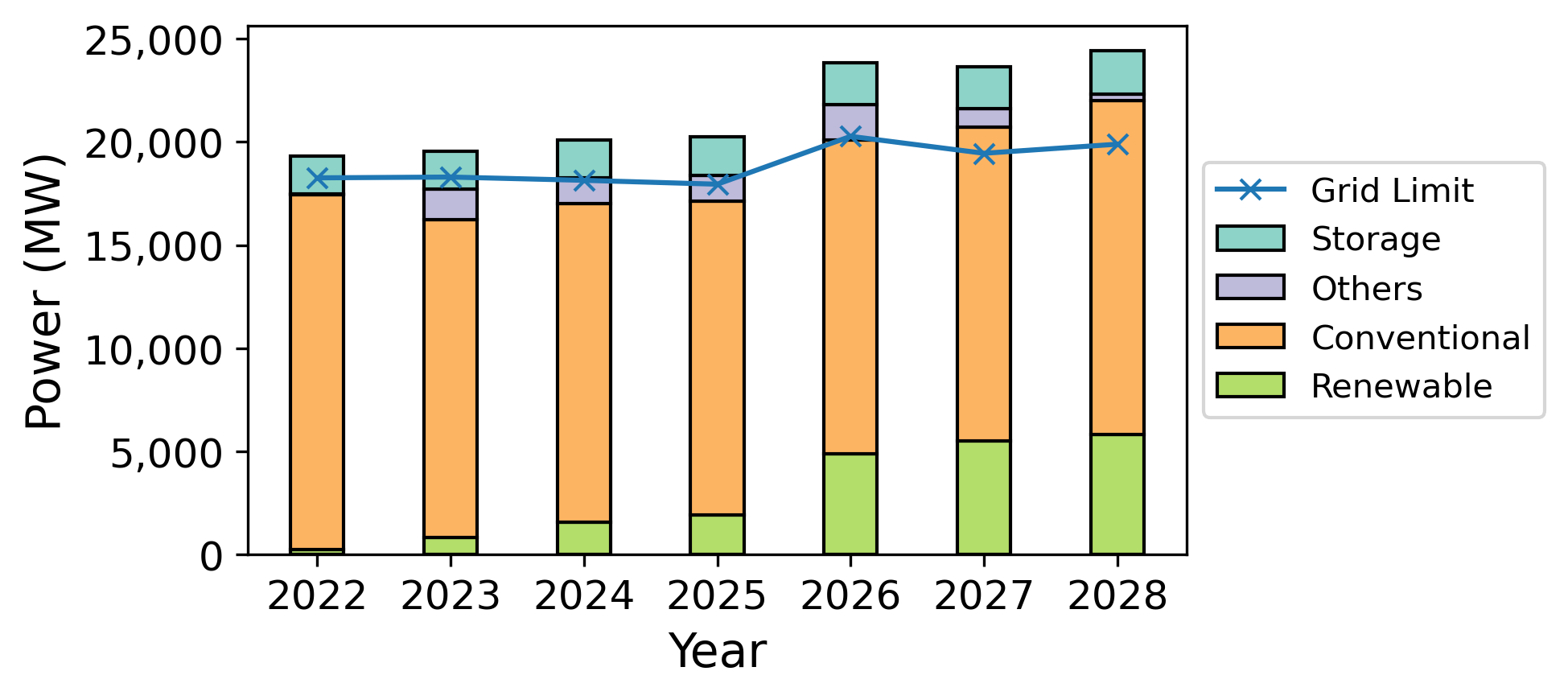}
    \caption{Dominion Reliable Load Limit and Resources.}
    \label{fig:dominion_resLimit}
\end{figure}

\begin{figure}[h]
    \centering 
    \includegraphics[width=\columnwidth]{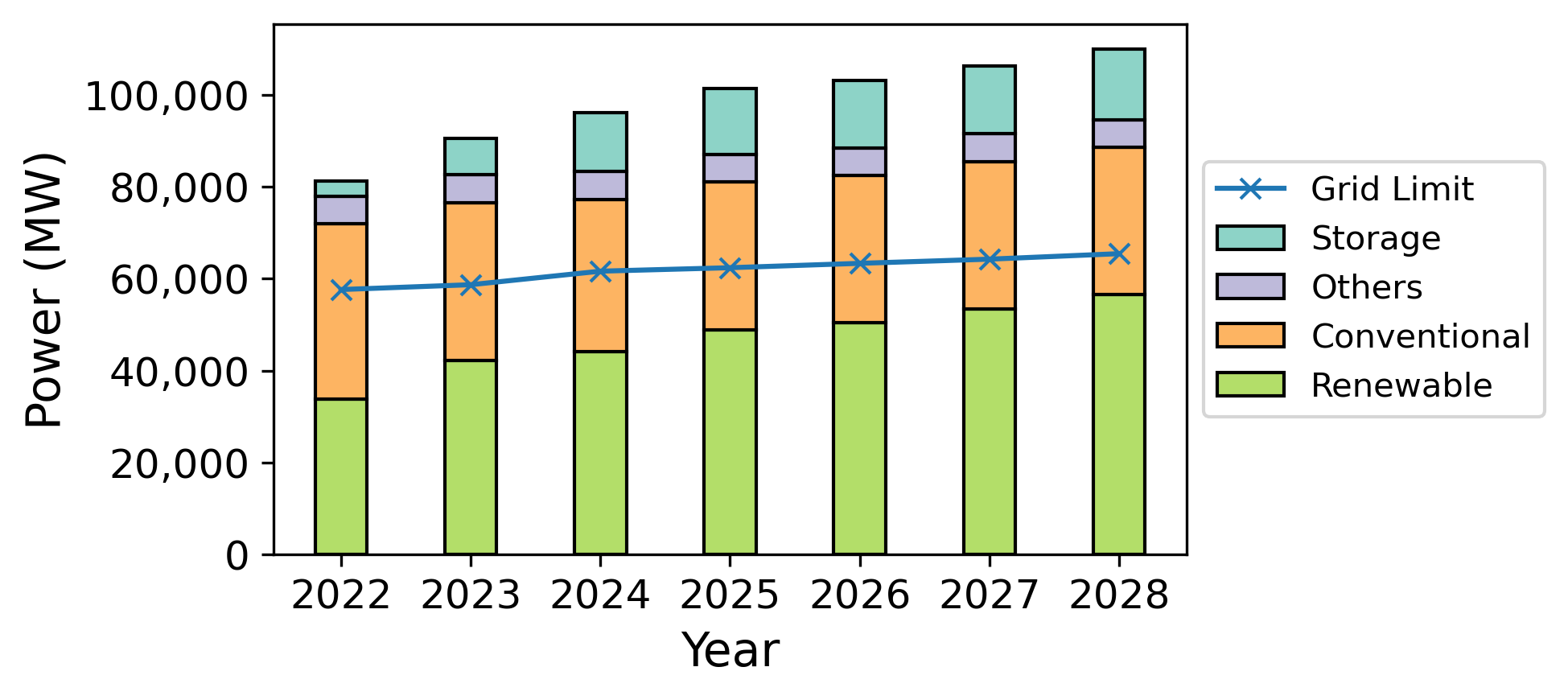}
    \caption{CAISO Reliable Load Limit and Resources.}
    \label{fig:caiso_resLimit}
\end{figure}

\begin{figure}[h]
    \centering 
    \includegraphics[width=\columnwidth]{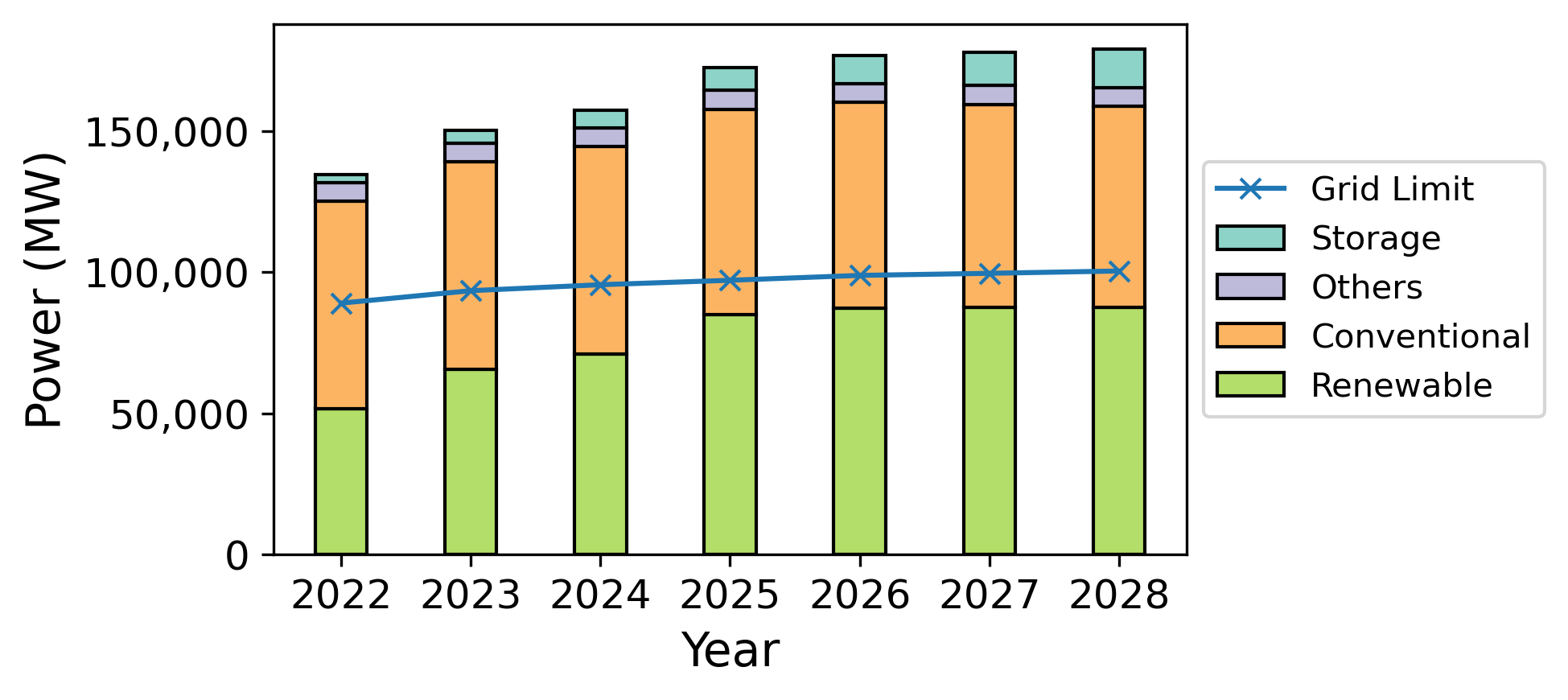}
    \caption{ERCOT Reliable Load Limit and Resources.}
    \label{fig:ercot_resLimit}
\end{figure}

\begin{figure}[h]
    \centering 
    \includegraphics[width=\columnwidth]{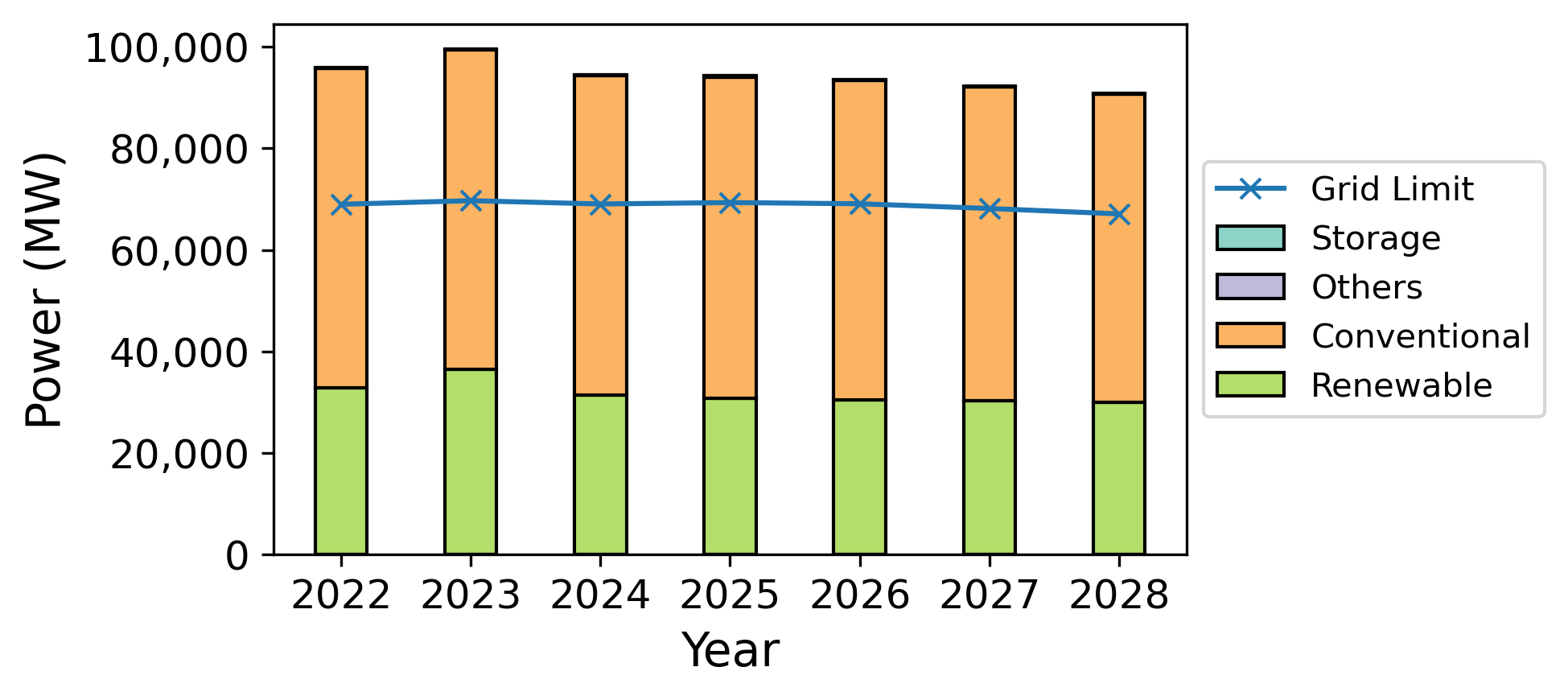}
    \caption{SPP Reliable Load Limit and Resources.}
    \label{fig:spp_resLimit}
\end{figure}

Figure \ref{fig:dominion_resLimit} to \ref{fig:spp_resLimit} depict the grid resource plan of Dominion, CAISO, ERCOT, and SPP. The diversity of resource mix and decarbonization progress can be clearly seen. As actual wind and solar generation are usually much less than their nameplate capacity (low capacity factors), the grid load limit is also much less than the total capacity in power grids with high fraction of renewable generation. For each time interval in a year, we scale 2022's renewable generation using the ratio of future capacity to 2022 capacity to model future renewable variation. We also separate non-DC load from load trace and scale it using the grid forecast to model future load variation.

Grid resource adequacy assessment considers the uncertainty of available resources. A conventional generator is not always available because of forced outages, maintenance, etc. Energy storage only shifts power, so how much it can discharge during grid peak hours depends how it is operated. Therefore, these resources are de-rated in adequacy assessment. De-rating factors in different grids are listed in Tabel \ref{tab:deratingFactors}. We apply CAISO's 0.73 energy storage de-rating factor to other grids as we think it's good practice and mainly depends on grid regulation that others can also adopt, but note that this could make the assessment results for ERCOT and EirGrid more optimistic than the reality. For the other resources, we directly use the de-rated values in grid reports.

\begin{table}[H]
    \centering
    \caption{De-rating Factors for Conventional Generation and Energy Storage of Different Grids}
    \begin{tabular}{lll}
        \toprule
         Grid&  Conventional& Energy Storage\\
         \midrule
         EirGrid& 0.75 \cite{eirgrid22report} & 0.73\\
         Dominion& 0.915 \cite{nercConvAvail} & 0.73\\
         CAISO& 0.915 \cite{nercConvAvail} & 0.73 \cite{caisoStorage}\\
         ERCOT& 0.89 \cite{ercot2023plan} & 0.73\\
         SPP& 0.894 \cite{spp2023plan} & 0.73\\
         \bottomrule
    \end{tabular}
    
    \label{tab:deratingFactors}
\end{table}

\section{Datacenter load growth forecast}
\label{appendix:loadGrowth}

In this section, we detail our model for projecting future demand on power grids from datacenter growth. Emerging generative AI applications are expected to accelerate the cloud-driven growth of datacenters. 

We project the incremental growth from AI based on NVIDIA's quarterly revenue forecast and history from its latest earnings call \cite{NVIDIAAugPressResease}. Under a flat annualization of quarterly revenue, and considering the price of A100 hardware (\$10,000) \cite{A100Price}, we derive the number of GPUs sold by NVIDIA. In accordance with the methodology in \cite{chien2023genai}, we then forecast the corresponding datacenter capacity backed by GPUs sold by NVIDIA, assuming 300W \cite{A100Specs} average incremental load (GPU at 50\% of TDP, server balance of 50\%, and datacenter PUE of 1.0). Using analyst estimates of NVIDIA's GPU market share of 68\% \cite{NVIDIAmarketShare}, we get AI's increase in the global datacenter capacity. By normalizing this increment to a median estimate of global datacenter capacity (33,000 MW) \cite{IEAdatacenterCapacity}, we arrive at the impact of GPU sales increase on datacenter growth---4.12\% increase in CAGR.

\begin{table}[h]
  \caption{AI Demand-driven Datacenter Growth Estimated from NVIDIA Revenue}
  \label{tab:load-model}
  \begin{tabular}{lll}
    \toprule
    Quarter Ending &Oct '23&April '23\\
    \midrule
    DC revenue (\$B) & 12& 4.28\\
    Annualized revenue(\$B) & 48& 17.12\\
    Number of GPUs sold (M)&4.8&1.712\\
    Resulting DC capacity (MW)&1,440&513.6\\
    Resulting DC capacity---Global (MW) &2,117.65&755.29\\
    Normalized to existing DC capacity &6.40\%&2.28\%\\
    \toprule
    Incremental Global DC capacity growth &4.12\%&\\
  \bottomrule
\end{tabular}
\end{table}

The incremental growth from AI is then added to the pre-AI datacenter growth. The average global datacenter growth is 7.21\% CAGR, as estimated from IEA's 2015--2022 data \cite{IEAdatacenterCapacity}. However, certain regions see higher growth because of suitable conditions for datacenters (e.g., networking, operation cost). We use grid-projected datacenter CAGR when available. For ERCOT and SPP, grid-projected DC forecasts are not available. We consider the ERCOT region (Texas) as a future hotspot for datacenters \cite{globalDC2023} that will see growth similar to Dominion's, and use the global average DC growth for SPP. We adjust the incremental growth using:
$$CAGR_{cloud+AI,r}=CAGR_{cloud,r}+\frac{CAGR_{cloud,r}}{CAGR_{cloud,avg}}\cdot \Delta CAGR_{AI,avg}$$
that assumes AI-demand incremental capacity in a region is proportional to the region's share in global datacenter capacity growth. The base Cloud growth rates and AI-accelerated growth are listed in Table ~\ref{tab:load-cagr}.

\begin{table}[h]
  \caption{Regional 2022 Datacenter Capacity and Estimated Growth Rate}
  \label{tab:load-cagr}
  \begin{tabular}{llll}
    \toprule
    Grid  & 2022 Capacity & Cloud Growth & Cloud+AI\\
    \midrule
    Dominion  &2767 MW \cite{dominionVA23report}& 15\% \cite{dominionVA23report} & 23.56\%\\
    ERCOT  &2332 MW \cite{usDCMap}& 15\% & 23.56\%\\
    EirGrid  &600 MW \cite{eirgrid22report}& 10.6\% \cite{eirgrid22report} & 16.65\%\\
    CAISO  &993 MW \cite{CADCGrowth} & 8.5\% \cite{CADCGrowth} & 13.35\%\\
    SPP  &1810 MW \cite{usDCMap}& 7.21\% & 11.33\%\\
  \bottomrule
\end{tabular}
\end{table}

\subsection{Revisiting AI Growth}
\label{appendix:loadGrowth_update}

In this section, we detail the updated linear DC load growth model described in Section \ref{sec:acceleratingGrowth}. We extend our AI load growth projections to consider NVIDIA's latest DC revenue for the quarters ending Oct '23 (actual, as opposed to the NVIDIA's forecast in section \ref{appendix:loadGrowth}), Jan '24 (actual) and April '24 (NVIDIA's forecast). Table \ref{tab:linear-load-model} shows published NVIDIA DC revenue and the growth in revenue between consecutive quarters. We calculate the average quarterly growth among the post-AI boom quarters, and extrapolate NVIDIA's DC revenue for subsequent quarters, assuming linear growth in revenue amounting to \$4.34B each quarter. The reminder of the projection methodology remains as described in section \ref{appendix:loadGrowth}.  

\begin{table}[h]
  \caption{NVIDIA's post-AI Boom quarterly DC revenue and growth from the previous quarter}
  \label{tab:linear-load-model}
  \begin{tabular}{lll}
    \toprule
    Quarter Ending & Revenue (\$B) & Growth (\$B)\\
    \midrule
    July '23 & 10.32 &  \\ 
    Oct '23 & 14.51 & 4.19\\
    Jan '24 & 18.40 & 3.89 \\ 
    April '24 & 23.34 &  4.94 \\ 
    \toprule
    Average quarterly revenue growth & &4.34 \\
  \bottomrule
\end{tabular}
\end{table}
\end{document}